\documentclass[12pt,hyper, notoc]{JHEP3}

\pdfoutput=1
\usepackage{amsmath,amssymb,calc}
\usepackage[bulletsep]{collref}
\usepackage{bbm}
\usepackage[pdftex]{graphicx}
\usepackage{verbatim}

\newcommand\be{\begin{equation}}
\newcommand\ee{\end{equation}}

\newcommand{\bea}{\begin{eqnarray}}
\newcommand{\eea}{\end{eqnarray}}
\newcommand{\half}{\frac{1}{2}}

\newcommand{\nn}{\nonumber}

\newcommand{\Tr} {{\textrm{Tr} }}
\newcommand{\STr}{{\textrm{STr} }}

\def\beq{\begin{equation}}
\def\eeq{\end{equation}}
\def\id{\protect{{1 \kern-.28em {\rm l}}}}

\def\unit{\relax{\rm 1\kern-.26em I}}

\title{Non Abelian Geometrical Tachyon}

\author{Vincenzo Cal\`o, Gianni Tallarita and Steven Thomas \\ \\ 
Queen Mary University of London \\
Center for Research in String Theory \\
Department of Physics, \\
Mile End Road, London, E1 4NS, UK. \\ 
Email: \email{V.Calo@qmul.ac.uk, G.Tallarita@qmul.ac.uk, S.Thomas@qmul.ac.uk}}

\abstract{We investigate the dynamics of a pair of coincident D5 branes in the background of $k$ NS5 branes. It has been proposed by Kutasov that the system with a single probing D-brane moving radially in this background  is dual to the tachyonic DBI action for a non-BPS Dp brane. We extend this proposal to the non-abelian case and find that the duality still holds provided one promotes the radial direction to a matrix valued field associated with a non-abelian  geometric tachyon  and a particular parametrization for the transverse scalar fields is chosen. The equations of motion of  a pair of coincident D5 branes moving in the NS5 background are determined.  Analytic and numerical solutions for the  pair are found in certain simplified cases in which the $U(2)$ symmetry is broken to $U(1) \times U(1)$  corresponding to a small transverse separation of the pair. For certain range of  parameters these solutions describe periodic motion of the centre of mass of the pair `bouncing off' a finite sized  throat whose minimum size is limited by the D5 branes separation.}

\preprint{QMUL-PH-}

\begin{document}
\tableofcontents
\section{Introduction}
Kutasov\cite{Kutasov:2004dj, Kutasov:2004ct} has presented intriguing links between systems of unstable D-branes and the DBI effective action of open string tachyon modes of non BPS D-branes. The former can be considered for example as a  probe BPS D-brane moving in a background geometry which breaks all remaining supersymmetry. An example of such a  geometric  background is that due to $k$ coincident NS5 branes \cite{Callan:1991at}. It then emerges that one can associate the radial motion of the probe brane in this background with that of  the  open string tachyon of non BPS D-branes. Such an association gives the former a geometrical interpretation, hence the notion of `geometric tachyons'. 

Kutasov's original model was further investigated and extended in \cite{Nakayama:2004yx, Sahakyan:2004cq, Kluson:2004xc, Kluson:2004yk, Chen:2004vw, Nakayama:2004ge, Thomas:2004cd, Thomas:2005am, Thomas:2005fw, Papantonopoulos:2006eg,  Sen:2007cz, Kluson:2007hb, Das:2008af}. Cosmological applications of geometrical tachyons were considered in \cite{Yavartanoo:2004wb, Thomas:2005fu, Ghodsi:2004wn, Panda:2005sg, Gumjudpai:2006hg, Panigrahi:2007sq, Panigrahi:2008kg} following Sen's original rolling tachyons ideas in \cite{Sen:2002nu}.

In this article we want to investigate what happens if we consider not just a single probe D-brane but rather a coincident pair of probe D-branes moving in the background of $k$ coincident NS5 branes. For $k$ large, this coincident pair of branes can still be regarded as probes in the sense that one may neglect the backreaction on the geometry to first approximation.

This paper is structured as follows: in Section 2 and Section 3 we consider different ansatz for the scalar fields which realise the map of the unstable D-brane system to one described by a non-abelian tachyonic mode. In section 4 we show the importance of a careful choice of definition for the harmonic function $H$ describing the NS5 branes background and we stress the differences between the matrix and function approach. In section 5 we make use of a symmetry breaking ansatz to expose a simplified version of the non-abelian system and give solutions for the equations of motion of the tachyonic field, these will be shown to reduce to the known single brane results in the abelian limit. Finally, in Section 6 we provide a small discussion of the results and suggest routes for further work.
\section{Multiple D-branes in the NS5-brane background }
Consider a stack of $k$ parallel NS5 branes in type II string theory, stretched in the directions $x^\mu=(t, x^1, \ldots, x^5)$, $\mu= 0, \ldots 5$, and localised in $x^m=(x^6, \ldots, x^9)$, $m=6,\ldots 9$. 
The background fields around $k$ parallel NS5 branes are 
\bea
&ds^2= G_{AB} dx^A dx^B = \eta_{\mu \nu} dx^\mu dx^\nu + \delta_{mn} H(x^m) dx^m dx^n &
\label{metric} \\
& e^{2\left( \Phi - \Phi_0 \right)} = H(x^m) &
\nn \\
& H_{mnp} = -\epsilon^q_{mnp}\partial_q \Phi & 
\eea
where the index $A = (\mu, m)$. The function $H(x^n)$ is the harmonic function describing $k$ five-branes, $H_{mnp}$ is the field strength of the Kalb-Ramond $B$-field and $\Phi$ is, as usual, the dilaton field. For coincident NS5 branes the harmonic function $H(x^n)$ reduces to
\be
H=1+\frac{kl_s^2}{r^2}
\ee
where $r=|\vec x|$ is the radial coordinate away from the five-branes in the transverse $R^4$ labeled by $(x^6,\cdots, x^9)$ and $l_s=\sqrt{\alpha'}$ is the string length. 

We are interested in the dynamics of two coincident BPS D5 branes in the background of the five-branes. We can label the world-volume coordinates of the D-branes by $\xi^\mu$, and by using reparametrization invariance on their world-volume we set $\xi^\mu = x^\mu$.

The low-energy dynamics of the D5-brane pair is described by a non-abelian U(2) gauge theory \cite{Myers:1999ps} (see also \cite{Tseytlin:1997csa}). The dynamics of the open string sector lightest degrees of freedom, namely the adjoint valued scalar fields $\left(X^6(\xi^\mu), \ldots, X^9(\xi^\mu) \right)$ which describe the position of the pair in the transverse directions $(x^6, \ldots, x^9)$, the non-abelian gauge field $A_{\mu}$ as well as the lightest degrees of freedom of the closed string sector, namely the metric $G_{AB}$, the dilaton $\phi$ and the Kalb-Ramond field $B_{AB}$ is governed by the non-abelian DBI action
\bea
\label{D5action}
S&=&\cr\nonumber
&-&T_5\int d^6x \, {\rm STr} \left( e^{-(\Phi-\Phi_0)} \sqrt{-det \left( P \left[ E_{\mu \nu}+E_{\mu m}\left(Q^{-1} - \delta \right)^{mn} E_{n\nu} \right] +\lambda F_{\mu \nu} \right) det(Q^m_n)} \right)
\eea
with
\be
\lambda = 2 \pi l_s^2, \quad E_{AB} = G_{AB} + B_{AB}, \quad \textrm{and} \quad Q^m_{\phantom{1}n} =\delta^m_{\phantom{1}n} + i \lambda [X^m, X^k] E_{kn}.
\ee
The field strength $F_{\mu\nu}=\partial_\mu A_\nu-\partial_\nu A_\mu-i[A_\mu,A_\nu]$, $P$ denotes the pullback to the brane world-volume and ${\rm STr}$ denotes the symmetrised trace.
%
%
%
%
\section{Fuzzy-sphere ansatz for the bulk scalars}
`Fuzzy sphere' configurations for the adjoint scalars $X^m$  in the previous non-abelian DBI action have been considered in the past  \cite{Myers:1999ps, Ramgoolam:2004gw,Thomas:2005fe,Thomas:2006ud} .  Let us generalise for the moment and consider the case of $N$ coincident D5-branes rather than just two and consider the following `fuzzy sphere' ansatz  for the transverse scalar fields:
\be
X^i = \hat{R}(x^\mu) \alpha^i \ , \quad i = 1, 2, 3,
\ee
where $\alpha^i$ give some $N\times N$ matrix representation of the $SU(2)$ algebra
\be
[\alpha_i, \alpha_j] = 2i \epsilon_{ijk} \alpha^k \ .
\ee
We define the physical radius of the 5-dimensional transverse space as
\be
R^2(x^\mu) = \frac{\lambda^2}{N} \sum_{i=1}^3 \, \textrm{Tr} \left[ X^i (x^\mu)^2 \right] = \lambda^2 C \hat{R}(x^\mu)^2
\ee
where $C$ is the Casimir of the particular representation of the generators under consideration, defined by the identity
\be
 \sum_{i=1}^3 \alpha^ i \alpha^i = C \id_{N}
\ee
Now, given this ansatz the DBI action becomes
\bea
S&=& -T_5\int d^6x \, {\rm STr} \left( \frac{1}{\sqrt{H}} \sqrt{1+\lambda^2 H \partial_a \hat{R} \partial^a \hat{R} \alpha^i \alpha^i} \sqrt{1+4\lambda^2 \hat{R}^4 H^2 \alpha^i \alpha^i} \right)
\label{fuzzyAction}
\eea
with
\be
H = 1 + \frac{k l_s^2}{\hat{R}^2}
\ee
where  it is understood that $H$ is a function of the physical radius $\hat{R}^2$, and not a matrix (this choice is investigated further in Section 4). Note that the symmetrized trace in the action ensures that one cannot simply replace all $\alpha^i\alpha^i$ by the Casimir $C$, there will be ordering issues which spoil this.

This action resembles a modified DBI action in flat background of $N$ non-BPS D5-branes proposed in \cite{Garousi:2008nj}, namely
\bea\label{DBI}
S_{DBI} &=& -T_5 \int d^6 x \ \STr \left( V(T_i T_i) \sqrt{1+\half \left[T_i, T_j \right]\left[T_j, T_i \right]} \right.
\nn \\ 
&& \phantom{1234567}
\left. \times \sqrt{-\textrm{det} \left(\eta_{ab}+\lambda \partial_a T_i \left(Q^{-1}\right)_{ij} \partial_b T_j \right)} \right)
\eea
where 
\be
Q_{ij} =  \id_N\delta_{ij} - i \left[T_i, T_j\right]
\ee
with $T_1=T\sigma_1$ and $T_2 = T\sigma_2$ and there is no sum over $i, j$.  Now we proceed to study two limits corresponding to regions of $\hat{R}$ space where the harmonic function $H$ takes two limiting forms, which correspond to the probe D-branes being close and far from the NS5's respectively. As will be shown, these limits need to be taken carefully in order to preserve forms of the DBI action which have sensible expandable forms.
\subsection{Large radius}

In this section we are looking for a limit of $\hat{R}$ space in which $H\rightarrow1$. Hence we need $R^2 >> kl_s^2$ but, to obtain an expandable DBI action which is crucial to performing calculations involving the Symmetrised Trace we must also have $\lambda R^2 =2\pi l_s^2 R^2 << 1$.

When this limit is taken the DBI action (\ref{fuzzyAction}) becomes:
\be
S= -T_5 \int d^6x STr\left(\frac{1}{\sqrt{H}}\sqrt{1+\lambda^2\partial_aR\partial^aR\alpha_m\alpha^m}\sqrt{1+4\lambda^2R^4\alpha_n\alpha^n} \right).
\ee	

Expanding both square roots and performing the symmetrised trace manually we find that this action is dual to

\be
S=-T_5 \int d^6x STr\left(V(T_iT_i)\sqrt{1+2T^4}\sqrt{1+\tilde{\lambda}\partial_aT\partial^aT\alpha_i\alpha^i}\right)
\ee

under the map $T^4= 2\lambda^2R^4CN$ (where $\tilde{\lambda}=\sqrt{\frac{N}{2C}}\lambda$ and $C$ is the Casimir of the $N$ dimensional representation generated by the $\alpha_i$) up to order $\lambda^2R^4$ which given the choice of limit means this term is small and higher order terms are progressively less important.
This is simply the large\footnote{large here is a slight misnomer, with the map used we still have $T^4<<1$, which is important for the latter action to be expandable} $T$ expansion of (\ref{DBI}), where to carry out this expansion one needs to be careful in the choice of $i=1,2$ as detailed in \cite{Garousi:2008nj}.

The potential takes the form:
\be
\frac{1}{T_5}V(T^2)=1-\frac{1}{2}\frac{2Ckl_s^2\tilde{\lambda}}{T^2}
\ee
which is simply the long range gravitational attraction between the D-branes and the five-branes.
\subsection{Small radius}
In this section we are looking for a region of $R$-space where 
\be
H \sim \frac{k l_s^2}{\hat{R}^2},
\ee
which is achieved for $R^2<kl_s^2$.  However we still want to have a DBI action which is expandable, hence we also need $R^2>\lambda^2 kl_s^2$, which is a sensible enough region provided $R^2$ is not too small compared to $kl_s^2$ originally (recalling that $\lambda=2\pi l_s^2$).

In this limit the DBI action becomes:
\bea\nonumber
S&=& -T_5\int d^6x \, {\rm STr} \left( \frac{\hat{R}}{\sqrt{k} l_s}  \sqrt{1+\lambda^2 \frac{k l_s^2}{\hat{R}^2 } \partial_a \hat{R} \partial^a \hat{R}\alpha^i \alpha^i} \sqrt{1+4\lambda^2 (k l_s^2)^2 \alpha^i \alpha^i} \right)
\eea
If we set
\be
T= \sqrt{k} l_s \ln \frac{\hat{R}}{\sqrt{k} l_s}
\ee
the previous action becomes,
\bea\nonumber
S&=& -T_5\int d^6x \, {\rm STr} \left( e^\frac{T}{\sqrt{k} l_s} \sqrt{1+4\lambda^2 (k l_s^2)^2 \alpha^i \alpha^i}\sqrt{1+\lambda^2 \partial_a T \partial^a T\alpha_i\alpha^i} \right)
\eea
which is the tachyon-DBI action with tachyon potential which is corrected from the usual $e^\frac{T}{\sqrt{k}l_s}$ by terms which are derived by expanding the action, taking the symmetrised trace and matching terms order by order.
%
%
\section{Commutative Ansatz}
Here we shall consider a different ansatz to that in the previous section. Inspired by \cite{Kutasov:2004dj}, where purely radial fluctuations of the fields on the branes give a geometrical description of a dual tachyonic system, we re-write the non-abelian  action in terms of a radial ``direction'' defined as $X^mX_m=R^2$, and we parametrize the scalar fields as 
\be
X^m = f^m(\theta, \phi, \chi) \tilde{R}
\label{commutingAns}
\ee
where $f^m$ are angular functions with $f^m f_m = 1$ and $R$ is an adjoint valued $U(2)$ matrix which we rewrite as a linear combination of $U(2)$ adjoint matrices $\alpha_a$ in the following way
\be
\tilde{R} = \tilde{R}_a(\xi) \alpha^{a}
\ee
where we have also included the $U(1)$ field $R_0$ and defined $\alpha_0 = \id_2$. With this parametrization it is clear that the commutator of the scalar fields vanishes $[X^m, X^n]=0$, in particular, $Q^m_n=\delta^m_n$ . Thus in contrast to the fuzzy sphere ansatz of the previous section, one might call this a `commutative' ansatz.

 The action of the D5-brane pair becomes:
\be\label{Action1}
S=-T_5\int d^6 x \, \STr  \left( \frac{1}{\sqrt{H}}\sqrt{-det\left(\eta_{\mu \nu}+H D_\mu R^a  D_\nu R^b \alpha_a \alpha_b+F_{\mu\nu} \right)}\right)
\ee
where
\be
H = 1 + \frac{k l_s^2}{X^m X_m}
\ee
where again  it is understood that $H$ is an $N \times N $ matrix, and
\be
X^m X_m = (R_0\alpha_0)^2 +2 R_0 R^i\alpha_0 \alpha_i + \left( R_i \alpha_i \right)^2
\ee
This action resembles that of two non-BPS D5-branes proposed in \cite{Garousi:2000tr} in the case of vanishing transverse scalar fields:
\be\label{Action2}
S= - T_5 \int d^{6}x \ \STr \left( V(T) \sqrt{-\textrm{det} \left(\eta_{\alpha \beta} + \lambda D_\mu T D_\nu T + \lambda F_{\mu \nu}\right)}\right).  
\ee
\subsection{Small radii limit}
In the limit in which  $R_0 \sim R_1 \sim R_2 \sim R_3 \sim 0$ the action (\ref{Action1}) reduces to
\bea
S&=& \cr\nonumber
&&-T_5\int d^6 x \, \STr  \frac{1}{\sqrt{H}} \sqrt{1+ k l_s^2 \frac{ \partial_\mu R_0 \partial^\mu R_0 \alpha_0^2 +2 \partial_\mu R_0 \partial^\mu R^i \alpha_i + \partial_\mu R^i \partial^\mu R^i \alpha_i \alpha_i  }{ (R_0\alpha_0)^2 +2 R_0 R^i\alpha_0 \alpha_i + \left( R_i \alpha_i \right)^2 }}
\eea
If we set
\be\label{maptt}
T = \sqrt{k} l_s \ln \left( R^m \alpha_m \right) =  \ln R
\ee
the above action becomes
\be
S=-\int d^6 x \, \STr \left( V(T)\sqrt{1+\partial_\mu T \partial^\mu T}\right)
\ee
and in this limit the potential is given by
\be
V(T) = \frac{T_5} {\sqrt{k} l_s}  e^{\frac{T}{ \sqrt{k} l_s}  }
\ee
If we define $T = T^m \alpha_m$ then in order to obtain an explicit expression for the different components $T^m$ of the tachyon matrix we would expand (\ref{maptt}) and match order by order each $T^m$ components on the l.h.s. with the respective $\alpha_m$ component on the r.h.s.\newline

There is an important point which must be noted here. The map \ref{maptt} is non-linear and hence to show that the duality truly holds one needs to show that this non-linearity is consistent in the symmetrisation procedure. The result quoted above for the dual action is true only under a symmetrisation with respect to the original field $R$, and not the new field $T$, which is what it would have to be in order for it to be the Tachyon DBI action. In order to show that this action is dual even under the symmetrisation procedure one needs to expand the original action in terms of $R$, perform the symmetrisation and then match order by order under a linear map for $T$. For this case this is a hard task due to the difficult powers of $\alpha_i\alpha_j$ appearing in the expansion and hence we will not show this and aim to investigate it further in future work.\newline

Note that this form of the map allows one to map the fully non-abelian actions, including the covariant derivatives. In particular, using eq. (\ref{maptt}), under the $\STr$ we have that
\bea
D_\alpha R&=&\partial_\alpha R + i[A_\alpha,R]\\
&=&\sqrt{k}l_s e^\frac{T}{\sqrt{k}l_s} \partial_\alpha T+i\sqrt{k}l_s \ e^\frac{T}{\sqrt{k}l_s}[A_\alpha,T] =  \sqrt{k}l_s\ e^\frac{T}{\sqrt{k}l_s} D_\alpha T
\eea
where in the second line we used the usual fact that $[f(R),\sigma_a]=f'(R)[R,\sigma_a]$ for $f(R)$ a continuous power series function of a matrix $R=R^a\sigma_a$. This means that
\be
\frac{1}{R^2} D_\alpha R D^\alpha R= D_\alpha T D^\alpha T +\frac{\sqrt{k}l_s}{R^2} [\exp{\frac{T}{\sqrt{k}l_s}},D_\alpha T].
\ee
The symmetrized trace  $\STr$ in the action will ensure that the commutator vanishes everywhere, so the non-abelian map including the covariant derivatives is realised in this limit.
\subsection{Large radii limit}
In the other case, namely when $R_0 \sim R_1 \sim R_2 \sim R_3 \rightarrow \infty$ the map is realized if we set
\be
T = R
\ee
and it is trivial to map the components of $T$ with those of $R$.
In this case the tachyon potential becomes
\be
V(T) = \frac{T_5}{\sqrt{1+\frac{k l_s^2}{T^2}}} \sim T_5 \left(1 -\frac{1}{2} \frac{k l_s^2}{T^2} \right)
\ee
which is the long-range gravitational attraction between multiple D5-branes and the NS5 branes.

In this limit it is trivial to map the covariant derivatives in the action, one simply has $D_\alpha T= D_\alpha R$ and also note that there are now no symmetrisation issues in matching the actions.
\subsection{General solution}
Given the ansatz in eq. (\ref{commutingAns}), we would like to show that one can find a general map for all values of $R$ between the two actions (\ref{Action1}) and (\ref{Action2}).  In \cite{Kutasov:2004dj}, it was shown that for the case of a single probing D-brane (where now $R$ is a function rather than a matrix) one could map the two systems by finding an analytical solution to the following differential equation:
\be
\frac{dT}{dR}=\sqrt{H(R)} 
\ee 
and by identifying the tachyon potential with the harmonic function $H$ as follows:
\be
V(T)=\frac{T_5}{\sqrt{H(R)}} \ .
\ee
In the small and large $R$ limits the map gave useful insight into the dynamics of the probing brane and provided useful information regarding rolling tachyonic solutions \cite{Kutasov:2004dj} and the nature of unstable D-brane systems. In the non-abelian case, the general requirement to realise the map is
\bea\label{STR}
\STr \left( H D_\mu R D_\nu R\right)&=& \STr \left( \lambda D_\mu T D_\nu T  \right)
\eea
When the system is promoted to a non-abelian one such a map is still possible. However, one needs to be careful with the choice of definition of $H$. One possibility is that $H$ can be thought of as a matrix, in which case $H(RR)$ depends in a general way on the matrix product of $R$, or we could understand $H$ to depend on $R$ via the non-abelian distance  $H(Tr(R^2))$ so that $H$ is a function and not a matrix. We will see through the rest of the paper that the choice is important. Careful string scattering calculations should reveal the true form of the Harmonic function appearing in the non-abelian DBI action and we think that once these calculations are performed the functional form of $H$ will
be obtained. However, being unaware of this result in the present literature, we decided to pursue both routes and obtain significantly differing results.
\subsubsection{H function}
Consider the case where $H$ is chosen to be a function. We will show here that analytical solutions for $T(R^a)$ still exist and furthermore that they yield the expected single brane results of \cite{Kutasov:2004dj} in the abelian limit. We consider the simplifying case where the gauge fields are turned off.

 In this case we define a physical radius as
\be
R^2 = \sum_{m=1}^3 \frac{1}{N} \Tr X^m X^m = \frac{1}{N} \Tr \tilde{R}^2 = R_0^2 + R_1^2 + R_2^2 + R_3^2
\label{physradius2}
\ee
With this choice one has\footnote{Notice that another ansatz which makes $H$ a function is $H\left(X^m X^m \right) \sim \Tr H\left(X^m X^m \right)$. This ansatz would lead to different results from those we find below and we do not pursue this approach any further.}
\be
H\left(X^m X^m \right) = 1 + \frac{k l_s^2}{\textrm{Tr} \ X^m X^m} = 1 + \frac{k l_s^2} {R_0^2 + R_1^2 + R_2^2 + R_3^2}
\ee
In this case we can solve the full map analytically. For every value of $R$ we need to solve
\be
\partial_\mu T =\sqrt{H(R^2)} \partial_\mu R \ .
\ee
If we write $T=T^m \alpha_m$ then for each $m = 0, \ldots 3$ we have to solve
\be
\partial_\mu T^m =\sqrt{ 1 + \frac{k l_s^2} {R_0^2 + R_1^2 + R_2^2 + R_3^2}
} \partial_\mu R^m.
\label{map}
\ee
In the abelian case with a single D5-brane we would find the solution
\be
\int \sqrt{1+\frac{k l_s^2}{R^2}} dR = \sqrt{k l_s^2 + R^2} + \half \ln \frac{\sqrt{k l_s^2 + R^2}-\sqrt{k}l_s}{\sqrt{k l_s^2 + R^2}+ \sqrt{k}l_s} = T^{kut}(R)
\ee
where $T^{kut} $ refers to the tachyon field of the single probe brane case of \cite{Kutasov:2004dj}.
By contrast, in the non abelian case we have to solve, for example, for the $m=0$ component
\be
\int \sqrt{1+\frac{k l_s^2}{R_0^2+d^2}} dR_0 = -i \sqrt{d^2+k l_s^2} E\left(i \sinh
   ^{-1}\left(\frac{R_0}{d}\right),\frac{d^2}{d^2+k l_s^2}\right)
\ee
where we define $d^2= R^2_1 + R^2_2 + R^2_3$ for simplicity and  $E(z,\omega) $ is the incomplete elliptic integral of the second kind. Although it is trivial to take the limit in which $d\rightarrow 0$ on the l.h.s., one has to take care with this limit  on the r.h.s. due to divergences appearing in the argument of the elliptic integral .
In order to explore the differences between the abelian and non-abelian case  it is instructive to expand the explicit expression for the integrand  on the l.h.s. in the limit in which $R^2_1 + R^2_2 + R^2_3 \ll R^2_0$
Then we obtain:
\bea
T_0 &= & T^{kut}(R_0) + \frac{1}{4} \frac{d^2}{R^2_0} \left(\sqrt{k l_s^2+R_0^2}- \frac{R_0^2\textrm{sinh}^{-1}
   \left(\frac{\sqrt{k} l_s}{R_0}\right)}{\sqrt{k} l_s}\right) \cr\nonumber
 &+& 
{\cal O}\left( \left( \frac{d^2}{R^2_0}\right)^4 \right),
\eea
the second term here denotes the non-abelian corrections to the abelian result.

For reference we write below the full solution for all $m$
\bea
T_m&=& \cr\nonumber
&&c(R^{j \neq m})-i \sqrt{kl_s^2+R^2 - \tilde{R}_m^2 } \text{E}\left[i \ \textrm{sinh}^{-1}\left[\tilde{R}^m \sqrt{\frac{1}{R^2- \tilde{R}_m^2}}\right],\frac{R^2- \tilde{R}_m^2}{kl_s^2+R^2- \tilde{R}_m^2}\right]
\eea
for $m = 0, 1, 2, 3$.
where $R^2 = {R^2_0 +R^2_1 + R^2_2 + R^2_3 }$ and $\tilde{R}_m$ corresponds to the component of $T_m$ one wants to solve for and $c(R^{j \neq m})$ is an integration constant.

\subsubsection{H Matrix}

In this case we would like to solve the map (\ref{STR}) where $H$ is in general a non-diagonal matrix. The map is non-trivial (in the case where no particular limit for $R$ is taken) unless $H$ is diagonalised, this can be achieved by choosing an a priori diagonal ansatz for $R$. In this case the full $U(2)$ symmetry of the problem would be broken to $U(1)\otimes U(1)$. To illustrate this take $R=R^0\sigma_0+R^3\sigma_3$, then 

\be
dT=\left(\begin{array}{rr}\sqrt{1-\frac{kl_s^2}{(R^0+R^3)^2}}&0\\0&\sqrt{1-\frac{kl_s^2}{(R^0-R^3)^2}}\end{array}\right)\times\left(\begin{array}{rr}dR^0+dR^3&0\\0&dR^0-dR^3\end{array}\right)
\ee

and substituting for $R_+=R^0+R^3, R_-=R^0-R^3,T_+=T^0+T^3$ and $T_-=T^0-T^3$ then one arrives at the map
\bea
dT_+&=&\sqrt{1-\frac{kl_s^2}{{R_+}^2}}dR_+\\
d T_-&=&\sqrt{1-\frac{kl_s^2}{{R_-}^2}}dR_-
\eea
which has as solutions two copies of the solution found in \cite{Kutasov:2004dj}. In particular, the action 
\be\label{action1}
S=-T_5 \int d^4x {\rm STr} \frac{1}{\sqrt{H}}\left(\sqrt{1-H\partial_\alpha R \partial^\alpha R}\right)
\ee
becomes
\be
S_T=-T_5 \int d^4x \left(e^{{\frac{T_+}{\sqrt{k}l_s}}}\sqrt{1-\partial_\alpha {T_+}\partial^\alpha {T_+}}+e^{{\frac{T_-}{\sqrt{k}l_s}}}\sqrt{1-\partial_\alpha{T_-}\partial^\alpha{T_-}}\right)
\ee
which is the $U(1)\otimes U(1)$ symmetric double copy of the single brane case. This is to be expected from a diagonal ansatz, the D-brane probes effectively separate and have independent single probe dynamics.

We believe that general results for arbitrary components of $H$ can be found but we will not pursue this any further as the calculations become very involved. However, as shown above one can indeed find maps for the $H$ matrix case by taking the limits of large/small $R$ first.
\section{Dynamics}
In the case where $H$ is regarded as a diagonal matrix  we have seen the effective action is just the direct sum of two independent actions each describing  the dynamics of a  single probe D5-brane,  which has already been investigated in \cite{Kutasov:2004dj}. Regarding $H$ as a function of the non-abelian distance defined in eq. (\ref{commutingAns}) produces a dynamical system where there is a non-trivial interaction between the  probe branes if we choose to separate them (which breaks $U(2) \rightarrow U(1) \times U(1) $) . Such an interaction vanishes in the flat space limit, as one would expect because then the probe branes are fully BPS and no force exists between them whether separated or coincident.

We take a symmetrical parametrization of the scalar fields, and demand that they depend only on time $t$ via
\be
X^m(t) = f^m(\theta, \phi, \chi) R(t) \ .
\ee
Starting from the following action\footnote{We set again the gauge fields to zero manually}
\be\label{action2}
S=-T_5\int d^6 x \, \STr  \left( \frac{1}{\sqrt{H}}\sqrt{-det\left(\eta_{\mu \nu}+H \partial_\mu R  \partial_\nu R\right)}\right) \ ,
\ee
we make a diagonal ansatz for the scalar field $R$
\be\label{diagonal}
R=R_0 \sigma_0 + R_3 \sigma_3
\ee
and finally we set
\bea
\phi&=&R_0+R_3\nonumber\\
\chi&=&R_0-R_3
\eea
One finds the action (\ref{action2}) reduces to 
\be
S=T_5\int d^6x \frac{1}{\sqrt{H}}\left(\sqrt{1-H\dot{\phi}^2}+\sqrt{1-H\dot{\chi}^2}\right)
\ee
where the harmonic function $H$ is now given by
\be
H=1+\frac{kl_s^2}{\chi^2+\phi^2}
\ee
The equation of motion that follows in the limit in which $\phi \sim \chi \ll k l_s^2$ is
\be
\frac{\phi}{k l_s^2 \sqrt{\frac{\chi^2+\phi^2}{k l_s^2}-\dot{\chi}^2}}+\frac{\phi}{k l_s^2 \sqrt{\frac{\chi^2+\phi^2}{k l_s^2}-\dot{\phi}^2}}+\frac{\ddot{\phi}}{\sqrt{\frac{\chi^2+\phi^2}{k l_s^2}-\dot{\phi}^2}}-\frac{\dot{\phi} \left(\chi \dot{\chi}+\dot{\phi} \left(\phi-k l_s^2 \ddot{\phi}\right)\right)}{k l_s^2 \left(\frac{\chi^2+\phi^2}{k l_s^2}-\dot{\phi}^2\right)^{3/2}}=0
\ee
with an analogous one in which $\chi$ and $\phi$ are interchanged. Exact solutions to these equations are hard to find but one can consider the conservation of the energy which results in a simpler first order differential equation.
The energy $E$ of the system is defined as
\be
E = P_{\phi} \dot{\phi} +  P_{\chi} \dot{\chi} - \mathcal{L}
\ee
and we investigate the following ansatz\footnote{by a slight abuse of notation we have used the notation $R_0$ though it is not strictly exactly the same as the  quantity  occurring in eq. (\ref{diagonal}).}
\bea
\label{simpleansatz}
\phi&=&\frac{1}{2}(R_0+C)\\
\chi&=&\frac{1}{2}(R_0-C)
\eea
where $C$ is a constant. In the small $R_0$ limit the conservation of the energy gives
\be\label{Energy}
\dot{R_0}^2=\frac{2\left(C^2+R_0^2\right)}{k l_s^2}-\frac{4T_5^2\left(C^2+R_0^2\right)^2}{E^2\left(k l_s^2\right)^2}. 
\ee
%
%
By imposing reality of the solution one obtains an important inequality
\be
\frac{2kl_s^2}{R_0^2+C^2} \geq \frac{4T_5^2}{E^2} -1,
\ee
we see that there are solutions at a critical energy $E_{crit}=2T_5$ which can escape to infinity.

The energy equation (\ref{Energy}) has analytical solutions for $C$ non-zero %
\be\label{jacobi}
R_0=\pm i C \text{JacobiSN}\left[\sqrt{\frac{2}{kl_s^2}} \sqrt{-1+2 \frac{T_5^2}{E^2} C^2} t \mp i \sqrt{-1+2 \frac{T_5^2}{E^2} C^2} c_1,\frac{2T_5^2 C^2}{-E^2+2T_5^2 C^2}\right]
\ee
where $c_1$ is an integration constant. In Figure 1 we present a plot of this solution  (with given choice of signs ) for certain values of the parameters $E, k, l_s,  c_1$ and the parameter $C=0,0.01,0.1$. They all correspond to the regime where the throat approximation to $H$ is valid. The case $C=0$ corresponds to the abelian case where $R_0(t)$ describes motion which is isomorphic to that of a single probe brane in an infinite throat, with energy less than the critical energy required to escape to infinity \cite{Kutasov:2004dj}.  What is particularly interesting in the case where
$C\neq 0$ is that the solutions appear to bounce, at least if we identify the solutions with negative values of $R_0$ as  separated probe branes moving up the throat. Looking at the harmonic function $H$ it is clear that in the case $C\neq 0$, the geometry seen by the probes is one of a finite cutoff throat, with $C$ acting as a cutoff parameter. So the resulting centre of mass dynamics of the separated probe pair is equivalent to a single probe brane moving in a cutoff throat background. 

In this interpretation, the sub-critical energy probe falls down the throat but then reflects off the  boundary and back up the throat reaching a certain maximum distance,  the motion being repeated forever.  It's clear from the plots that the period of oscillation increases with decreasing $C$. This makes sense as in the limit $C\rightarrow 0$ we recover the solution found in  \cite{Kutasov:2004dj} which does not oscillate (at least not in coordinate time  $t$ ) but corresponds to an infalling probe brane taking infinite coordinate time to reach the throat bottom. By patching the two solutions (which differ by a minus sign) together in the regions where $R_0(t)$ is negative one finds an explicit change of sign in the velocity $\dot{R}_0(t)$ of the branes as they reach the throat cutoff, as is expected from a perfectly elastic bouncing solution.
\begin{figure}[ht]
\centering
\includegraphics[width=0.8\linewidth]{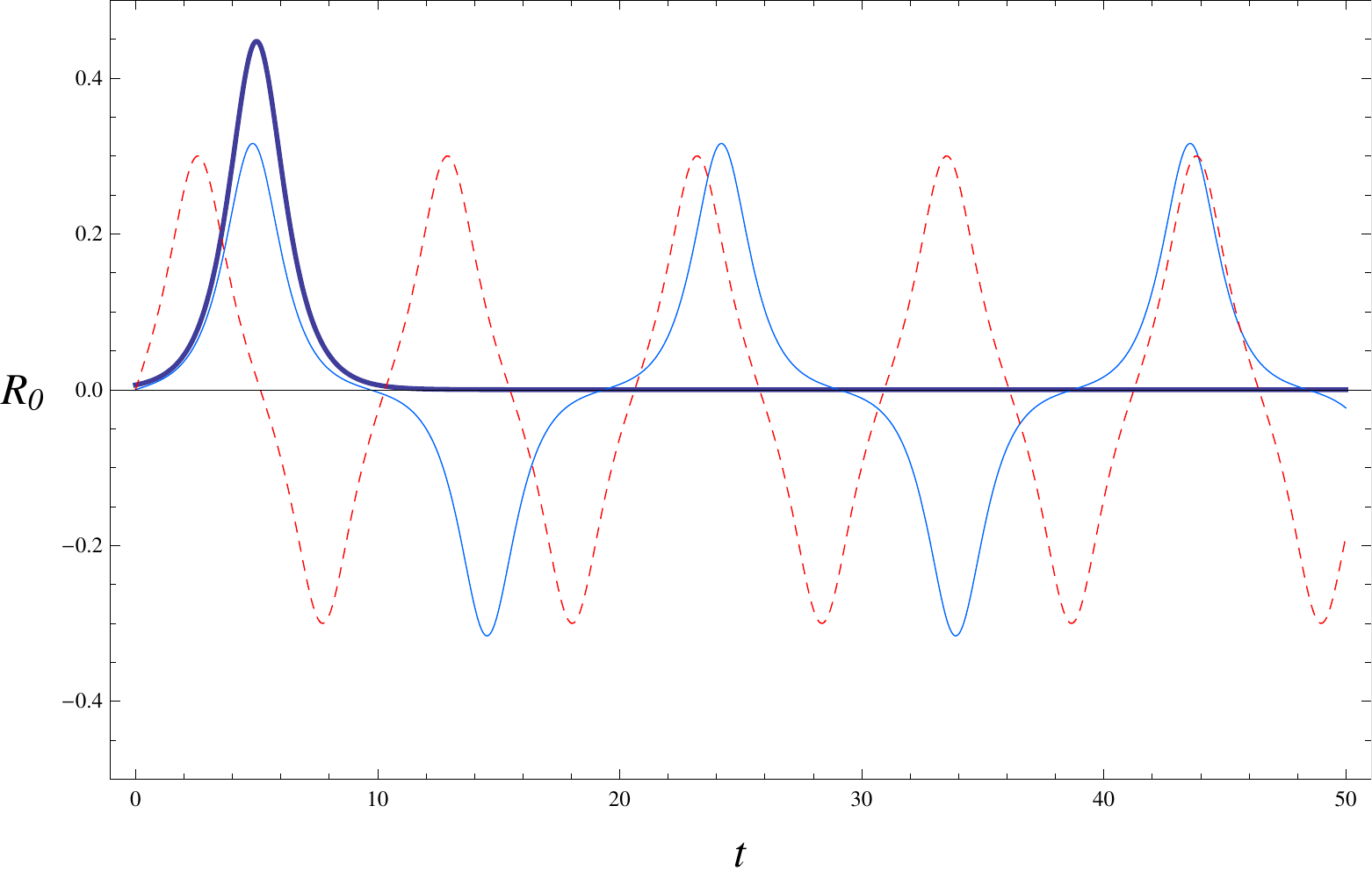}
\caption{Plots of $R_0$ vs $t$. The bold curve has $C=0$ and corresponds to the abelian case. The continuous regular curve corresponds to $C=0.01$, and finally the dashed curve to $C=0.1$. In all  cases we have chosen $\frac{2T_5^2}{E^2}=10$ , $c_1=0$ and $kl_s^2=1$}\label{Figure1}
\end{figure}
We now investigate the behaviour of the string coupling with time using the relation
\be \label{dilaton}
e^{2\phi}=g_sH(Tr(RR))
\ee
In Figure 2 we show a plot of the effective string coupling using the solution (\ref{jacobi}), valid in the throat approximation.  The thick curve corresponds to the abelian case $C=0$ and shows, as one expects, a rapidly increasing effective string coupling as the probe falls down the infinite throat. Thus after some time $t=t_{max}$  the solution is no longer within the perturbative string approximation. As argued in \cite{Kutasov:2004dj}, the value of $t_{max}$ depends analytically on the energy $E$ of the probe and there is an energy  `window' $g_s T_5 \ll  E \ll T_5 $ for which the probe brane moves in the throat and remains within perturbation theory. 

By contrast , the case where $C \neq 0$ (regular and dashed curves in Figure 2), we see the effective coupling as oscillating in time as the probes oscillate in the throat. By choosing the value of $E$ and/or $C$ it is possible to control the motion such that the string coupling is always in the perturbative regime and for the probes to remain in the throat region for all time.

Due to the complexity of the solution (\ref{jacobi})  one cannot derive a simple expression for a bound on the energy and/or $C$ in order for the above to hold, even for small $C$. Instead one has to use the full expression for the JacobiSN function for $C\neq 0$ and thus we are limited to numerical plots as in Figure 2.
\begin{figure}[ht]
\centering
\includegraphics[width=0.8\linewidth]{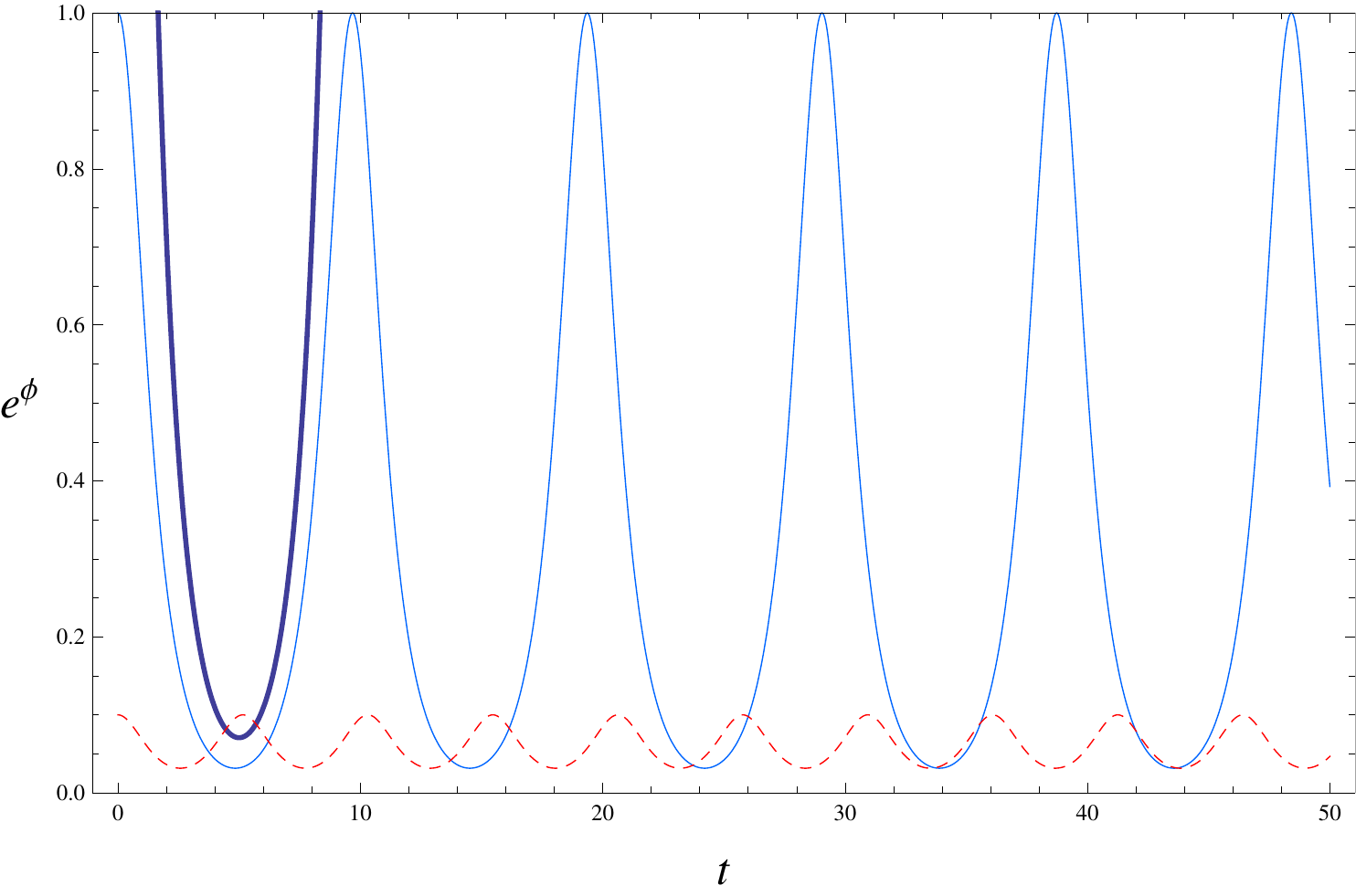}
\caption{Plots of the effective string coupling $e^{\phi}$ vs $t$. 
 The bold curve has $C=0$ and corresponds to the abelian case. The continuous regular curve corresponds to $C=0.01$, and finally the dashed curve to $C=0.1$. In all plots  we have chosen $\frac{2T_5^2}{E^2}=10$ , $c_1=0$ and $kl_s^2=1$. The value of  $g_s=0.0001$.} 
\label{Figure2}
\end{figure}
\section{Discussion}
In this article we have attempted to generalize the notion of Kutasov's geometric interpretation of the open string tachyon, in the scenario where a D-brane is moving in a background geometry of $k$ NS5 branes that render the system non-BPS \cite{Kutasov:2004dj}.
The generalization we investigated considered a pair of coincident probe D5-branes moving in this background instead of a single probe D5 brane discussed in \cite{Kutasov:2004dj}. The single real geometric tachyon field that appears in the single probe case
is, in the simplest scenario, related to purely transverse radial motion of the  probe. This system is abelian in that there is a $U(1)$  gauge theory on the probe brane world volume. 

When we consider the case where, for example, one has as a probe two coincident D5-branes, then the situation becomes more subtle. Firstly the probe world-volume now supports non-abelian $U(2)$ gauge fields and secondly, as is well known, the coordinates transverse to this probe stack become matrix valued.   This latter phenomenon raises the question of how one interprets the geometrical quantities such as the harmonic function $H$ sourced by the NS5 branes. In one interpretation, we can define a notion of non-abelian distance in the transverse matrix geometry via the quantity $Tr(X^mX^m ) $ where $X^m$ are the matrix valued transverse coordinates. Then $H(X^m)$ can be thought of  as a function  via  $H= H(Tr(X^mX^m))$. Another  possible interpretation is that $H$ becomes a matrix through its dependence on $X^m$.

Both definitions seem to give rise to well defined actions since ultimately the lagrangians are matrix valued objects in each case and Str is taken over all free gauge indices. However as we have shown, the resulting  definition of the matrix valued geometric tachyon field and the resulting dynamics is different in the two interpretations.

As an illustration of this we saw that in the case where $H$ is treated as a function of non-abelian distance defined above, the 
tachyon map can be found exactly and in the limit where the $U(2)$ adjoint valued radial coordinate $R$ is dominated by the terms proportional to the $2 \times 2$ identity matrix, we recovered the single probe brane tachyon map of Kutasov.

On the other hand, a general solution  for the tachyon map in the case of $ H $ being a matrix is very complicated and its explicit form is not known. However we found that at least in the symmetry breaking case where $U(2) \rightarrow U(1) \times U(1) $ the system reduces to two non-interacting copies of single geometrical tachyon fields. 
By contrast, the same  $U(2)$ breaking configuration of the probe stack, in the case where $H$ is a function and not a matrix, yields a dynamical system involving two coupled geometric tachyon fields. 

In this case we found analytic expressions for  homogeneous time dependent solutions  at least in the situation where  we consider only diagonal degrees of freedom in the non-abelian tachyon field, which corresponds to $U(2)$ symmetry breaking.
Interestingly we found oscillating or `bouncing' solutions in this case where the separation parameter between the two D5 probes acting as an effective cutoff on the NS5 brane infinite throat. 

It would be very interesting to find (even numerically) dynamical  solutions which  involve the  full non-abelian degrees of freedom in the $U(2)$ valued tachyon field in the action eq (5.2) including non-vanishing gauge fields. 
Another extension could be to look at different arrangements of background NS5 branes other than the point like ones considered in this paper. For example one can also consider the  k NS5 branes  arranged around a ring of finite radius. This is a known supergravity solution and the  corresponding metric and harmonic function are known \cite{Sfetsos:1998xd,Sfetsos:1999pq} . This would extend to the non abelian case the results found in \cite{Thomas:2004cd}.

\providecommand{\href}[2]{#2}\begingroup\raggedright\endgroup

\end{document}